\pdfoutput=1
\documentclass[a4paper,fleqn]{cas-dc}

\usepackage[authoryear,longnamesfirst]{natbib}
\usepackage{graphicx}
\usepackage{listings}
\usepackage{float}
\usepackage{pifont}
\usepackage{wasysym}
\usepackage{xcolor}

\lstdefinestyle{py}{
  language=Python,
  basicstyle=\ttfamily\footnotesize,
  keywordstyle=\color{blue}\bfseries,
  commentstyle=\color{gray}\itshape,
  stringstyle=\color{orange},
  showstringspaces=false,
  breaklines=true,
  frame=single,
  rulecolor=\color{gray},
  aboveskip=3pt,
  belowskip=3pt,
}

\lstdefinestyle{diff}{
  language=Python,
  basicstyle=\ttfamily\footnotesize,
  morecomment=[l][\color{green!45!black}\bfseries]{+},
  morecomment=[l][\color{red!55!black}\bfseries]{-},
  showstringspaces=false,
  breaklines=true,
  frame=single,
  rulecolor=\color{gray},
  aboveskip=3pt,
  belowskip=3pt,
}

\begin{document}

\let\WriteBookmarks\relax
\shortauthors{Gustavo~Pinto}
\shorttitle{Access More Than Reasoning}

\title [mode = title]{Where LLM Agents Get Stuck Debugging Databases: Access More Than Reasoning}

\author[1]{Gustavo Pinto}[orcid=0000-0001-7598-2799]
\ead{gpinto@ufpa.br}
\cormark[1]
\cortext[1]{Corresponding author}
\affiliation[1]{organization={Federal University of Par\'a (UFPA)},
            city={Bel\'em},
            state={Par\'a},
            country={Brazil}}

\begin{abstract}
Large language model (LLM) coding agents are increasingly asked not only to write code but to
debug running systems from live runtime data. We report a field observation of one such agent
(Claude Code) as it diagnosed eight database incidents in a production message-dispatch engine.
In every incident whose root cause depended on runtime concurrency, the agent announced a cause,
shipped a fix, and later walked it back, and each wrong guess cost a deploy-and-observe cycle.
Mapping each incident to what the agent lacked, we find that access to production state was the
most pervasive bottleneck. The missing access included live lock views, untruncated deadlock
reports, read-only execution, and an explicit environment context. An access gap was present in
all five concurrency-dependent incidents: alone in one, and in four combined with a reasoning
error such as anchoring on an early cause, incomplete exploration of the configuration, or a
misreading of the code's transaction semantics. The three incidents that did not depend on
concurrency needed nothing beyond what the agent already had. Within the incidents, the agent's
wrong hypotheses were consistent with the evidence it could see; they were not fabrications. We
argue that the next step for production-debugging agents is a forensics layer that gives the
agent the access it lacks inside existing security controls. Such a layer would combine a safe
interrogation API, a lock-and-deadlock flight recorder, and a production replay harness.
\end{abstract}


\maketitle

\section{Introduction}\label{sec:introduction}

Large language model (LLM) coding agents are widely used to write and review code
\citep{sweagent,watanabe2026agentic}. A harder use is now emerging: teams give these agents live
runtime data, such as logs, database state, and transient failures, and ask them to debug
production systems. Unlike code completion, this requires acting on a live system under time
pressure, with real consequences. Industry has started building LLM agents for incident
management \citep{rcacopilot,roy2024rca,rcagent}, and the database community has proposed agents
specialized in database diagnosis \citep{dbot,panda}.

Among the systems these agents are asked to debug, databases are a particularly difficult
class. Failures such as lock contention, deadlocks, and race conditions emerge from the
interaction of concurrent workloads, and they are short-lived: by the time the agent looks at
the logs, the lock may already be gone. The agent must therefore diagnose the failure from
after-the-fact evidence rather than from a reproducible state. Getting the diagnosis wrong is
costly. A missed root cause means a deploy that does not fix the problem, or a second incident,
and each redeploy runs through a slow delivery pipeline (Section~\ref{sec:discussion}). The
cost is also human: while the failure persists, the developers on call are under pressure to
restore service, and every fix that does not hold sends them back to waiting on the pipeline.
An agent that converges in one attempt rather than several is worth a great deal.

We know surprisingly little about where such agents get stuck. Most of the research focuses on
how well agents write code \citep{swebench,agentless}, not on how well they debug running
systems. Where operations agents are evaluated, it is mostly on benchmarks that inject faults
into controlled environments with complete telemetry \citep{aiopslab,itbench,openrca}.
As a result, the natural assumption that the limit is the model's reasoning has gone largely
untested in the field.

The assumption matters because it determines where effort goes. If the agent's mistakes are
reasoning failures, or hallucinations in the sense of confident fabrications \citep{zhang2025hallucination},
then the answer is a more capable, better-grounded model. If instead the limit is access to
logs, to the database, or to a way of reproducing the failure, then investment in larger models
addresses the wrong problem. Access is also hard to grant. Companies wrap production in layers
of security, such as VPNs, identity management, and read-only proxies, that an agent cannot
cross directly. In practice, a human engineer still does the manual work the agent would
otherwise automate: searching logs, running read-only SQL against production, and reading raw
output to reconstruct what happened.

This paper studies that question in the field. We follow an LLM agent (Claude Code) as it
diagnosed and fixed a series of real database incidents in a production message-dispatch engine
at a large technology company. We combine the agent's session transcripts, which contain its
private reasoning and its public answers, with the repository's commit history, which
independently records each fix that was made and each fix that was undone. Our research question
is:

\begin{center}
\fbox{\parbox{0.85\linewidth}{\centering
\textbf{RQ.} What limits an LLM coding agent's diagnosis of production database
incidents: its reasoning or its access?}}
\end{center}

Our answer is a qualified one: mostly access. An access gap was present in all five incidents
whose root cause depended on runtime concurrency. In one of them it was the only limit. In the
other four it combined with a reasoning error, and in each of those the error was corrected only
through production evidence that the agent could not obtain directly. Within the incidents, the
agent's wrong guesses were reasonable inferences from incomplete information rather than
fabrications.

This paper makes the following contributions:
\begin{itemize}
\item to our knowledge, the first field report of an LLM agent debugging database incidents in
a live system, using the agent's own transcripts and the commit history as primary evidence,
including a record of how it converged by serial self-correction at the cost of a deploy cycle
per wrong guess;
\item a \emph{reasoning} $\times$ \emph{access} matrix, derived from the eight incidents, that
separates the gaps in the agent's access from its reasoning errors and shows where the two
interact;
\item a concrete agenda for tool builders, made of a safe database forensics API, a
lock-and-deadlock flight recorder, and a production replay harness, that restores the agent's
access without lowering the security controls already in place.
\end{itemize}

\section{Background}\label{sec:background}

\subsection{The System and Its Workload}\label{sec:system}

The system under study is a message-dispatch engine (WhatsApp, SMS, and e-mail) for B2B
marketing campaigns, operated by a large technology company. The stack is Python and SQLAlchemy
(asyncpg) behind PgBouncer, on Kubernetes, with Datadog observability. The engine follows a
three-stage pipeline:

\begin{itemize}
\item \textbf{Compile.} A trigger (a scheduler tick, a manual action, or an event) invokes the
journey compiler. The compiler resolves the audience from a customer data platform (CDP),
materializes per-customer attributes from an entity--attribute--value (EAV) store of about
38.5~million rows across 369 dynamic columns, and applies capping and routing rules.

\item \textbf{Orchestration.} The engine selects eligible customers, enforcing one message per
customer per day, a phone-number cascade, and coverage constraints. It materializes them as rows
in the dispatch-queue table, which is the single source of truth. This stage has produced
3{,}089 journey executions.

\item \textbf{Dispatch.} A worker claims queue rows in batches of one hundred under advisory
locks and sends each message through the appropriate delivery provider. Webhooks then record the
delivery status.
\end{itemize}

\begin{figure*}[ht]
\centering
\includegraphics[width=\textwidth]{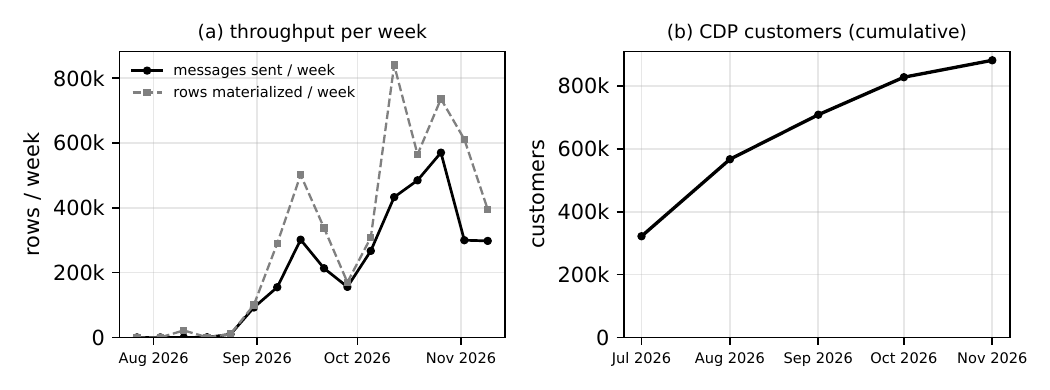}
\caption{Dispatch workload over time: (a) messages sent and queue rows materialized per week;
(b) cumulative distinct customers in the CDP.}
\label{fig:workload}
\end{figure*}

The workload explains why the locking layer is under pressure. The CDP's distinct-customer base
grew from roughly 330{,}000 in July to about 587{,}000 by November
(Figure~\ref{fig:workload}b). Over its lifetime the dispatch queue has accumulated about
4.9~million rows and sent about 3.3~million messages. A single journey can materialize hundreds
of thousands of queue rows; the largest execution held 331{,}000, and up to 22{,}000 rows were
observed pending at once. Dispatch load is concentrated in the morning, with the heaviest
sending window between 09:00 and 12:00 local time; in October, that three-hour window
accumulated roughly 600{,}000 messages over the month. Figure~\ref{fig:workload} shows the
ramp-up over time.

\subsection{Development Context}\label{sec:dev}

\begin{figure*}[ht]
\centering
\includegraphics[width=\textwidth]{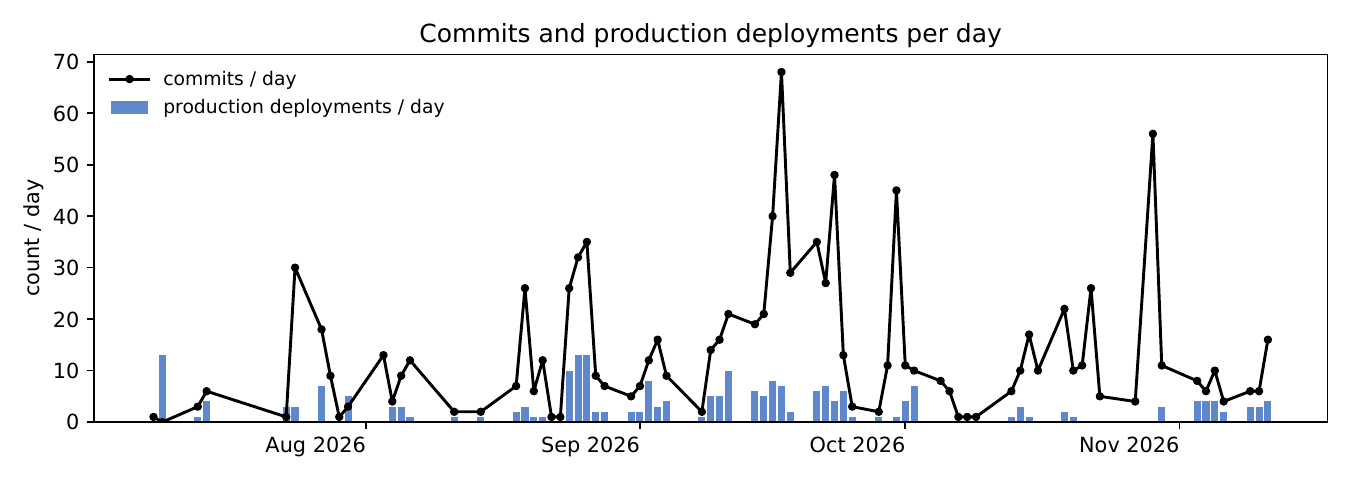}
\caption{Commits and production deployments per day over the project lifetime (July--November
2026): blue bars are production deployments, the black line with markers is commits.}
\label{fig:deploys}
\end{figure*}

The repository holds roughly 406{,}000 lines of source code (Python, TypeScript, and SQL),
about 182{,}000 of them tests, with 84\% line coverage. Over four months it accumulated
1{,}019 commits by five developers, of which 293 (29\%) carry an AI co-author trailer: 289 name
Claude Code and four name Cursor. The trailer is used by only two of the five developers, so it
understates the agent's involvement. Production was deployed on most business days, with 232
production deployments over 126 days (Figure~\ref{fig:deploys}).

The codebase did not start from scratch. It was first built with an AI coding agent on the
Replit platform and was imported into the company's infrastructure in a single commit of about
241{,}000 lines in July 2026. Many defects of that version surfaced only under production load
and real data. We return to this origin as a threat to validity in
Section~\ref{sec:limitations}.

The engineer instructs Claude Code through natural-language prompts in a spec-first and
test-first style: each non-trivial change starts with a short specification, is implemented
against a failing test, and is verified with a regression test. The agent runs with elevated
permissions (\texttt{bypassPermissions}) and commits under the engineer's identity. A persistent
memory directory (\texttt{.memory/}) records decisions, lessons, and, most relevant here,
refuted hypotheses, so that a fresh session can resume without re-litigating settled questions.
Every change passes through a merge request that a second developer must approve.

\section{Method}\label{sec:method}

This study is a field observation conducted through participant observation \citep{ko2017}. The
author was employed at the company under study while the incidents occurred and directed the
agent as part of ordinary work. The analysis is retrospective: it reconstructs the agent's
behavior from artifacts produced during that work, not from a controlled experiment. The company
requested anonymity, so we refer to it only as ``a large technology company.''

The unit of analysis is the incident. We study eight database incidents that occurred over a
ten-day window, each a distinct failure the agent was asked to diagnose and fix. We also draw on
sessions from the preceding two weeks when they bear on the same failure surface. We lack both a
controlled comparison, since we cannot vary the agent's access or its reasoning, and ground truth
about what would have happened otherwise. We therefore triangulate three independent artifacts:
session transcripts, commit history, and deployment and workload records. Each captures a
different facet of the agent's behavior, and we treat agreement across them as the evidence.

The study spans 27 agent sessions and about 410 hours of cumulative session time, during which
the agent issued about 3{,}000 tool calls. The session time is the sum, per session, of the last
event timestamp minus the first; sessions stayed open across working days rather than running
continuously. Roughly two thirds of the tool calls were shell commands; the rest were file
edits, task tracking, and platform API calls. Claude Code's model calls go through the company's
internal generative-AI platform, which exposes several models; the user selects the model in the
tool's configuration, and the session logs record the model that served each call. Most calls
were served by an open-weights model (DeepSeek, about 79\%), which the company encourages for
cost, and a minority by a frontier model (Claude Opus, about 17\%). We did not attribute
individual hypotheses to models, so we make no claim about whether a different model would have
avoided a given false start (Section~\ref{sec:limitations}).

\subsection{Study Setting and Access}\label{sec:access}

The agent's access was asymmetric: broad in the development environment and narrow in
production. In development it had everything a developer has:

\begin{itemize}
\item \textbf{Repository:} read and write access to the local clone, and access to the remote
repository through a GitLab MCP server, including merge requests and history;
\item \textbf{Local execution:} the logs of the application running locally;
\item \textbf{Tests:} the output of the unit tests and of the integration tests, which run
against a local database.
\end{itemize}

This access let the agent read any line of code and reproduce any failure that a single local
process with a local database can produce. The incidents we study are the ones it cannot: their
failures emerge only under production's concurrent load and data. For this class of runtime
failures, access to the code is necessary but not sufficient. The code says what \emph{can}
happen; resolving the incident requires knowing what \emph{did} happen, and that knowledge lives
in production.

Into production, by contrast, the agent has three partial channels, all incomplete:

\begin{itemize}
\item \textbf{Logs via skills:} log search over Datadog and Kubernetes. Indexed datasets require
browser authentication, which the agent does not have.

\item \textbf{Read-only database:} the agent writes the SQL, but a human runs it through a
read-only production-database proxy, which cannot execute, scale, or restart anything. A human
could therefore run a \texttt{pg\_locks} query for the agent, but each query took a round trip
through a person and returned a snapshot taken after the agent asked, by which time a transient
lock could already be gone.

\item \textbf{kubectl:} cluster access, but without \emph{exec} into the database, without the
ability to restart or scale pods, and with an environment context that must be switched
manually.
\end{itemize}

\subsection{Data Sources and Analysis}\label{sec:evidence-collection}

We draw on three artifacts. First, the agent's session transcripts (JSONL), which record both
its private reasoning (\texttt{thinking}) and its public answers (\texttt{text}). This
trajectory-level view follows recent studies of coding-agent behavior \citep{bouzenia2025trajectories}, but
applies it to a production setting rather than a benchmark. Second, the repository's commit
history, which records each fix the agent shipped and each fix it later corrected or reverted.
Third, deployment and workload records: production deployments from the platform's API, and
read-only database queries covering the customer base and the hourly dispatch volume.

\emph{Incident selection.} We included every database incident in the window that the team
worked on with the agent. Five of the eight began with a false start, two did not, and in one
the first fix only mitigated the symptom.

We analyzed the data in two passes:

\begin{itemize}
\item \textbf{Qualitative.} We read the transcripts incident by incident and reconstructed the
sequence of hypotheses and fixes. We coded each hypothesis as \emph{held} or \emph{retracted}
using two independent signals: the agent's own later statements (``The fix did NOT resolve the
slowness'') and the commit history (a later commit that corrects an earlier one). A \emph{false
start} is a hypothesis or fix that at least one of these signals marks as retracted. For each
retracted hypothesis we then coded its source using two operational criteria. A hypothesis is
\emph{evidence-consistent} if it agreed with everything the agent had observed up to that point
in the session; it is a \emph{fabrication} if it asserted a fact contradicted by evidence already
in the agent's context. We also recorded whether the information that eventually refuted it was
available to the agent (in the repository, the documentation, or its tools) or out of its reach.
The sessions are mostly in Portuguese, with some agent turns in English; quotations from
Portuguese turns are the author's translations, and all quotations are otherwise unedited except
that system- and company-identifying names are replaced with generic descriptors.

\item \textbf{Quantitative.} We measured, per incident, the number of fix commits, whether the
first attempt held, and the number of production deployments on the incident's day
(Table~\ref{tab:incidents}). Commits and lines of code come from \texttt{git}; code size at a
revision is the number of newline characters in all tracked \texttt{.py}, \texttt{.ts},
\texttt{.tsx}, and \texttt{.sql} files. Deployments come from the platform's deployment API,
filtered to production environments.
\end{itemize}

\section{Database Incidents and the Agent's False Starts}\label{sec:incidents}

Eight incidents occurred over ten days, and six of them converged on a single failure surface:
the locking and transactional-consistency layer of the dispatch engine.
Table~\ref{tab:incidents} lists them and quantifies the pattern they share: the agent was never
wrong in the end, but it was often wrong \emph{serially}: it announced a cause, shipped a fix,
and walked it back, at the cost of a deploy cycle per wrong guess. The repository records the
same churn from the outside: \texttt{dispatch\_service.py} alone was touched fourteen times. The
production deployment record shows the cost: on every incident day the team shipped between two
and four production deployments. We trace each incident, and the evidence of its false start,
below.

\begin{table*}[ht]
\caption{The eight incidents, their fix commits, the production deployments on the incident's
day, and whether the first attempt held.}
\label{tab:incidents}
\centering
\footnotesize
\begin{tabular}{l l l p{1.7in} c c c p{1.2in}}
\hline
\textbf{\#} & \textbf{Day} & \textbf{Incident} & \textbf{Description} & \textbf{Conc.} &
\textbf{Fixes} & \textbf{Deploys} & \textbf{First attempt held?} \\
\hline
\#1 & D2 & Memory exhaustion & OOM kills; stale read in the compile handler & --- & 2 & 2 & Partly---symptom only \\
\#2 & D3 & Summary lock & lock timeout; row lock held to commit & \lightning & 3 & 3 & No---failure continued \\
\#3 & D4 & Pair lock and ledger & unlock returned \emph{false}; one session per event & \lightning & 6 & 3 & No---corrected by next \\
\#4 & D5 & Trigger committed early & double-send risk; pause on bad phone & --- & 2 & 2 & Yes \\
\#5 & D8 & Monitor card & journey canceled by orchestration & --- & 1 & 1 & Yes \\
\#6 & D9 & Throughput & the ``tick'' hypothesis refuted & \lightning & 0 & 0 & No---six hypotheses refuted \\
\#7 & D10 & Race exposed & a purge $\times$ orchestration race & \lightning & 5 & 2 & No---wrong cause \\
\#8 & D10 & Deadlock retry & retry rolled back between attempts & \lightning & 3 & 1 & No---rollback was wrong \\
\hline
\end{tabular}

Note: \lightning{} marks the incidents whose root cause depended on runtime concurrency (locks,
races, or concurrent throughput); all five began with a false start. Day is relative to the
ten-day incident window (D1--D10). Fixes counts fix commits. Deploys counts the production
deployments that carried each incident's fixes. The throughput incident has no fix commit: its
fix was a configuration change, raising the retry-sweep cap from five to eight pairs per tick in
the deployment values, not a code change; its six hypotheses were refuted by measurement before
that change. On the shared day of \#7 and \#8, eight fix commits crossed two merge requests:
five belong to the race and three to the deadlock retry.
\end{table*}

\subsection{Memory exhaustion (---)}\label{sec:incident-memory}

The segmentation step loaded an entire audience into memory in one go, and the pod ran out of
memory and was killed by the operating system (\emph{OOMKilled}, exit 137), again and again,
across all five pods. At the same time, the compile error handler read an object after its
database transaction had already been rolled back, raising \texttt{MissingGreenlet}. The quick
fix was to throw resources at the problem: the pod's CPU request went from 50\,m to 200\,m, its
CPU limit from 500\,m to 1000\,m, and its memory limit from 2\,Gi to 4\,Gi. This fix stopped the
out-of-memory kills, so we code its first attempt as having held only in part. The underlying
cause, materializing the whole audience instead of streaming it, was left open, and removing the
memory pressure exposed the summary-lock timeout of incident \#2: ``The OOMKilled did not come
back, good. But the lock timeout persists.''

\subsection{Summary lock (\lightning)}\label{sec:incident-summary}

Two parts of the engine were fighting over the same database row: the dispatch summary tried to
write its result while holding a lock that its own next step then had to wait on, so it timed
out waiting on itself. The agent's first fix was an early commit: write the status sooner, so
the lock is released. That was the false start: the logs show ``\texttt{summary\_persist\_failed}
continues (12 in 6min) \ldots{} The fix did NOT resolve the slowness.'' Only a third commit,
which stopped the summary from holding that row lock at all, worked. Listing~\ref{fig:diff-summary}
shows the eventual fix: instead of \texttt{flush()}, which held the row lock until the final
commit, the code now \texttt{refresh()}es after the separate summary session commits.

\begin{lstlisting}[style=diff, captionpos=b, caption={The summary-lock fix, shown as a diff.}, label=fig:diff-summary]
-    execucao.exec_metadata = meta   # marks the object dirty
-    await self.db.flush()           # flush holds the row lock until commit
+    await self.db.refresh(execucao)  # re-read after the summary session commits
\end{lstlisting}

\subsection{Pair lock and ledger (\lightning)}\label{sec:incident-pair}

When the engine locked a pair of (journey, execution) to dispatch it, the unlock step silently
failed---it returned \texttt{false}, and no one checked the return value. The agent's first
guess was to start checking that boolean; the real fix was the opposite---stop closing the
transaction, so the lock stays on the same database connection under PgBouncer. The agent caught
its own mistake mid-investigation: ``But wait, line 1056 \emph{does} await
\texttt{self.db.commit()}, so there is a contradiction.'' A second problem sat in the same area:
the ledger opened one database session per event, about 600 round-trips per batch, adding write
contention; it was collapsed to one session per batch. Listing~\ref{fig:diff-pair} shows the
pair-lock fix: checking the boolean became keeping the transaction open.

\begin{lstlisting}[style=diff, captionpos=b, caption={The pair-lock fix, shown as a diff.}, label=fig:diff-pair]
-    res = await lock_conn.execute(unlock_sql)
-    released = bool(res.scalar())
+    if got_lock:
+        # do NOT rollback(): keep the PgBouncer
+        # connection pinned to this client
+        await self._dispatch_impl(...)
\end{lstlisting}

\subsection{Trigger committed early (---)}\label{sec:incident-trigger}

When a journey was triggered, the system saved (committed) the queue too early, before the
orchestration step got a chance to run. If that step then failed, the orphaned rows were swept
up and sent without any of the usual checks, risking the same message going out twice to one
customer. The fix was to hold everything in a savepoint instead of committing, so a failure
rolls the whole thing back rather than leaving the queue half-written.
Listing~\ref{fig:diff-trigger} shows the change: the per-journey \texttt{commit()} became a
\texttt{begin\_nested()} savepoint.

\begin{lstlisting}[style=diff, captionpos=b, caption={The trigger fix, shown as a diff.}, label=fig:diff-trigger]
-    await self.db.commit()
-    # queue persisted BEFORE orchestration
+    async with self.db.begin_nested():
+        compile_journey(journey)
+        run_for_day(journey)
+    # savepoint: one failure rolls back all
\end{lstlisting}

\subsection{Monitor card (---)}\label{sec:incident-monitor}

One journey simply vanished from the operator's screen. It had been canceled by the
orchestration because another journey (same segment, same product, lower priority) had been
paused since October; the execution closed with zero of 10{,}000 messages sent, and the card
disappeared without explanation. The monitor was changed to keep recently-finished journeys
visible for 24 hours and to show why they were blocked.

\subsection{Throughput (\lightning)}\label{sec:incident-throughput}

An investigation that ran across sessions from D3 to D9 was one long false start: the agent
proposed and discarded cause after cause before finding the real one. It began with the
preflight query, guessing the per-message checks were the bottleneck, until a measurement put
that at 0.34\,s for 100 rows, far too cheap to explain the lag. It then blamed the scheduler
tick, but raising the tick did not move throughput. Next the batch cap, but the effective rate
was two batches per pass, below the cap of three. Then line eligibility, but the executions were
terminating (11{,}476 sent, zero pending), so nothing was stuck. Then the volume throttle, but
its value was NULL on every active journey. It even shortened the orphan grace period, and
reverted it when it did not help. Only after all six did it find the real bottleneck: the retry
sweep's five-pairs-per-tick cap, which saturated during draining. The fix was a one-line
configuration change that raised the cap, not a code change. We count this incident as
concurrency-dependent because the bottleneck appeared only when many executions drained at once,
a state that exists only under production's concurrent load.

\subsection{Race exposed (\lightning)}\label{sec:incident-race}

An optimization that read a batch of rows up front accidentally widened the gap between reading
and writing, and another process (the compile purge) deleted rows in that gap, outside the lock.
When the orchestration then tried to update those rows, it found zero, and the whole tick
aborted before anything was sent. The agent first chased the wrong lead (``This is the smoking
gun \ldots{} the issue is \emph{[a frozen attribute store]}''), a real but historical problem
from late September, and even started in the wrong environment (``No, good catch. I was on
sandbox \ldots{} let me switch to the prod context''). The error it was chasing was this, raised
when the per-day write found the row already deleted:

\begin{quote}\footnotesize
\texttt{StaleDataError: UPDATE statement on table 'dispatch\_queue'\\
expected to update 1 row(s); 0 were matched}
\end{quote}

In plain terms: the code tried to update a row that another process had already deleted, and the
database library, expecting exactly one row, aborted the whole tick before any message could be
sent. Listing~\ref{fig:diff-race} shows the eventual fix.

\begin{lstlisting}[style=diff, captionpos=b, caption={The race fix, shown as a diff.}, label=fig:diff-race]
-    row = await self.db.get(DispatchQueue, row_id)
-    row.status = CANCELLED                 # row may be purged
+    stmt = (update(DispatchQueue)
+            .where(DispatchQueue.id == row_id,
+                   DispatchQueue.status == PENDING)
+            .values(status=CANCELLED))
+    await self.db.execute(stmt)            # 0 rows: no-op
\end{lstlisting}

\subsection{Deadlock retry (\lightning)}\label{sec:incident-retry}

The agent's own hotfix introduced a new bug. To make the orchestration survive a deadlock, it
added a retry that called \texttt{rollback()} between attempts, and its own comment claimed this
was safe (``rollback discards only the orchestration state''). It was not: the rollback wiped
the trigger's still-uncommitted work and expired every loaded object, so the next read raised
\texttt{MissingGreenlet}, the same bug the hotfix had just fixed in the dispatch path. The fix
was to retry inside a savepoint, never a rollback. Listing~\ref{fig:diff-retry} shows the change.

\begin{lstlisting}[style=diff, captionpos=b, caption={The retry fix, shown as a diff.}, label=fig:diff-retry]
-            await db.rollback()      # drops compiles, expires ORM objects
+            async with db.begin_nested():  # retry in a nested transaction
\end{lstlisting}

\subsection{Costly Convergence}\label{sec:costly-convergence}

Seven of the eight incidents were resolved by a fix that held, and production recovered
afterward. In the eighth (\#1), the fix resolved the symptom and the underlying cause was
deferred; the streaming rewrite of the segmentation step was not shipped during the study
window. Every final fix was shipped by the team, and in the incidents covered by the transcripts
it was the agent that wrote it. The pattern we report is therefore not one of failure but of
costly convergence: the agent reached the right diagnosis every time, but often only after one
or more wrong guesses. In our sample, the split follows concurrency. The two incidents with no
false start (\#4 and \#5) and the one whose first fix only mitigated the symptom (\#1) did not
depend on runtime concurrency. All five concurrency-dependent incidents began with a false
start; in them the root cause was hard to see without access to runtime state.

\subsection{Why Database Bugs Are the Hard Case}\label{sec:why-hard}

Most of these false starts are the cost of diagnosing from incomplete and transient signals, and
none of them is a hallucination in our coding. A hallucination is a fabrication: it asserts a
fact that the evidence already in view contradicts. A false start, as we use the term, is a
hypothesis consistent with everything the agent had seen that turns out to be wrong. The two
call for different remedies: hallucination calls for better grounding, and false starts call for
better access. This reading is consistent with evidence that current LLMs detect data races and
deadlocks reliably when the whole program is in view \citep{jain2026concurrency}. Our agent could also read
the whole program, in the sense that every file of the repository was available to it, but two
differences matter. First, those benchmark programs are small enough to fit in the model's
context, whereas our codebase has about 406{,}000 lines. The agent could read only the parts it
chose to search, so a relevant fact could sit in the repository and still go unread, as the
retry-sweep cap of incident \#6 shows. Second, and more important for concurrency, the code
describes which interleavings are possible, not which one occurred. That information lives in the
runtime state, such as the locks held, the rows deleted, and the transactions left open, and the
agent could not observe it.

Not every error fits this reading, however. In four incidents the information that refuted the
wrong hypothesis was within the agent's reach. In the pair lock (\#3), the agent's first guess
was contradicted by a line of code in the repository, and the agent caught the contradiction
itself mid-investigation. In the throughput incident (\#6), the retry-sweep cap that turned out
to be the bottleneck sat in the deployment values in the repository, and the agent explored six
other causes before looking there. In the race (\#7), the agent anchored on a real but historical
cause from late September. In the deadlock retry (\#8), the agent's own comment claimed that
\texttt{rollback()} was safe between attempts; SQLAlchemy's documented semantics say otherwise.
We code these as reasoning errors. In each case, though, an access gap was also present and
shaped the error or delayed its correction: the missing live lock view in \#3, the missing live
metrics in \#6, and the wrong environment in \#7. In \#8, the truncated deadlock report hid the
deadlock cycle, so the agent could not remove the deadlock's cause and chose to survive it with
a retry; the retry's rollback was the reasoning error. We therefore classify these four
incidents as mixed in Section~\ref{sec:matrix}. We also observed one case close to a
fabrication, outside the eight incidents: three days before the window (D-3), the agent reported
a fix as done that had never reached the code, and it caught the discrepancy itself in a later
turn (``The fix was NOT applied \ldots{} the bug is still in \texttt{app.py:51}''). When the
agent concluded that the preflight query was the bottleneck, it was not hallucinating; it was
reasoning from the data it could see. It could read the retry-sweep cap in the deployment
values, but it could not see that the cap was saturating during draining, which only live
throughput metrics would have shown. The agent cannot run queries against the production
database, so it must infer row-level state from logs: ``I do not have access to the database nor
\emph{exec} in prod (I tried, no permission).'' Locks and deadlocks are transient by nature: by
the time the agent looks, the lock is gone, and the deadlock detail arrives truncated (``the
DETAIL is STILL truncated \ldots{} at the SOURCE''):

\begin{quote}\footnotesize
\texttt{DeadlockDetectedError: deadlock detected\\
DETAIL: Process 3051747 waits for AccessExclusiveLock\\
on tuple (128477,20) of
relation 17528 \ldots\ [truncated]}
\end{quote}

A deadlock happens when two transactions each hold a lock the other needs. The error's
\texttt{DETAIL} is the lock graph. It names which transaction is waiting for which lock and
which transaction holds it, and it is precisely what identifies the deadlock cycle. Here the log
was cut off at the source, so the agent could see that one process was waiting for a lock on a
specific row, but not which process held that lock or what the rest of the cycle was. It knew a
deadlock had happened, but not who was involved in it, and without that the root cause is a
guess. The truncation itself is telling: the deadlock \texttt{DETAIL} is long, because it lists
every process, what each waits for, and what blocks each. The application's logging layer stores
the exception in a fixed-length field that cuts it off partway. The full detail existed in the
Postgres server's own log---which has \texttt{log\_lock\_waits} enabled and logs the complete
deadlock report---but the agent had access only to the application log. The information was not
absent; it was out of reach.

Even obtaining access cost a deploy. To see live lock state during the summary-lock incident,
the agent had to write a \texttt{pg\_locks} probe and ship it to production (``The
\texttt{pg\_locks} probe \ldots{} will show the \texttt{blocker\_query}''), turning a read into a
release. Each wrong hypothesis then costs a full deploy--observe cycle before it can be refuted,
and the indexed Datadog logs that would settle a diagnosis sit behind a login the agent cannot
pass (``requires login''). For a bug class whose evidence is momentary and whose reproduction
requires the concurrent load and external-system integration that a local replica cannot
provide, the agent has no shortcut: it must guess, ship, and watch.

\section{The Reasoning $\times$ Access Matrix}\label{sec:matrix}

For each incident we asked two questions. What did the agent lack in order to close the
diagnosis? Was that missing piece a property of the environment (access) or of the agent
(reasoning)? We coded an incident as access when the evidence that refuted the wrong hypothesis
was out of the agent's reach, as reasoning when it was available but went unused or was misread,
and as mixed when both applied. Incidents without a wrong hypothesis are coded sufficient.
Table~\ref{tab:matrix} summarizes the result.

\begin{table*}[ht]
\caption{What the agent lacked, by incident.}
\label{tab:matrix}
\centering
\footnotesize
\begin{tabular}{l l c p{1.5in} p{1.6in} l}
\hline
\textbf{\#} & \textbf{Incident} & \textbf{Conc.} & \textbf{Access gap} & \textbf{Reasoning error} & \textbf{Limit} \\
\hline
\#1 & Memory exhaustion & --- & --- & --- (cause identified; fix deferred) & Sufficient \\
\#2 & Summary lock & \lightning & Live \texttt{pg\_locks} view & --- & Access \\
\#3 & Pair lock and ledger & \lightning & Live \texttt{pg\_locks} view & First guess contradicted by the code & Mixed \\
\#4 & Trigger committed early & --- & --- & --- & Sufficient \\
\#5 & Monitor card & --- & --- & --- & Sufficient \\
\#6 & Throughput & \lightning & Live throughput metrics to test each hypothesis & Cap in deployment values not inspected & Mixed \\
\#7 & Race exposed & \lightning & Explicit environment context & Anchoring on a historical cause & Mixed \\
\#8 & Deadlock retry & \lightning & Full deadlock \texttt{DETAIL} (truncated) & Misread rollback semantics & Mixed \\
\hline
\end{tabular}
\end{table*}

The matrix splits cleanly along concurrency. The three incidents that did not depend on
concurrency needed nothing the agent did not already have; in \#1 the diagnosis was right and
only the full fix was deferred. Every one of the five concurrency-dependent incidents had an
access gap. In one (\#2) it was the only limit. In the other four it combined with a reasoning
error, and no incident was limited by reasoning alone.

The mixed rows deserve a careful reading, because they are where a more capable model is most
plausibly useful. In each of them, the reasoning error was retired by evidence rather than by
reflection: a measurement refuted the throughput hypotheses, the historical cause was set aside
once it was shown to date from September, and a renewed failure exposed the unsafe rollback.
Better access would have shortened each of these loops, because it would have exposed the error
before a deploy rather than after it. Whether a stronger model would have avoided the errors in
the first place is a question our data cannot answer (Section~\ref{sec:limitations}).

We therefore state our central finding with that qualification: in our observation, access to
production state was the most pervasive limit on the agent's diagnosis, and where reasoning
failed, it failed alongside an access gap that shaped the error or delayed its correction. The
implication runs against the assumption that the model is the ceiling. In this setting, the
instrumentation was at least as binding.

\section{Discussion}\label{sec:discussion}

\subsection{Summary of Findings}\label{sec:summary-findings}

The paper's main findings are threefold. First, the agent diagnosed by serial self-correction.
It announced a cause, shipped a fix, and walked it back, at the cost of a deploy cycle per wrong
guess. Across eight incidents it shipped twenty-two fix commits. All five concurrency-dependent
incidents began with a false start, and the throughput investigation alone burned six
hypotheses. Second, access was the most pervasive bottleneck. The reasoning $\times$ access
matrix shows an access gap in all five concurrency-dependent incidents, alone in one and combined
with a reasoning error in four, and no incident limited by reasoning alone. Third, this access is
hard to grant by design, because production sits behind layers of security. The agent ran mostly
on an efficient open-weights model, with a frontier model as a minority. Because we did not
attribute hypotheses to models, this mix neither supports nor undermines the access thesis; it
is an open question that a controlled replay could settle (Section~\ref{sec:conclusion}).

A corollary follows for the engineers who operate these agents. The false starts were caught
only because a human knew enough about concurrency to judge the hypothesis, demand the right
evidence (a live lock view, an untruncated deadlock report), and verify that the fix held. The
agent did not replace this expertise. It amplified its leverage, but only where a human could
tell when the agent was wrong.

\subsection{Implications for Tooling}\label{sec:tooling}

If access is the dominant bottleneck, the practical question for researchers and tool builders is
how to give an agent better access for production database debugging. Our evidence suggests five
concrete directions. The first three are the components of the forensics layer named in the
abstract; the last two concern the agent's workflow.

One caveat frames what follows: this access cannot simply be granted. The target is a production
database, and opening it to an agent is not a matter of toggling a permission. The environment
sits behind a VPN and an identity-management layer, and every access to a service is brokered by
a local access-control application that decides who may reach what. Any tooling must therefore
live inside these layers, as a read-only, audited forensics surface rather than raw database
credentials. This constraint is, in fact, why a purpose-built interrogation layer, and not a
blanket grant of access, is the right shape for the problem.

\textbf{A safe interrogation layer for production databases.} The agent's largest gap is that it
cannot ask the production database anything beyond a read-only projection: ``I do not have access
to the database nor \emph{exec} in prod.'' What it needs is a forensics API: a read-only surface
exposing \texttt{pg\_locks}, \texttt{pg\_stat\_activity}, query plans, and row-level state under
the safety guarantees that already protect production, but rich enough to answer ``which
statement holds this lock?'' This is the runtime counterpart of what SWE-agent calls the
agent--computer interface \citep{sweagent}, and of Panda's grounding and affordance components
\citep{panda}, but it is exposed to an application-side agent under production security
constraints rather than to a DBA tool. Standardizing such an API for agents is a concrete
deliverable.

\textbf{Persistent capture of transient concurrency.} Locks and deadlocks are momentary; the
agent consistently arrived after the fact and found nothing (``the DETAIL is STILL truncated
\ldots{} at the SOURCE''). A ``flight recorder'' that continuously logs lock waits and full
deadlock reports, instead of a log line that truncates at the interesting part, would turn an
ephemeral failure into a queryable artifact.

\textbf{Reproducing production's emergent behavior.} The agent can replicate data volume locally,
but not the concurrent workload and external-system integrations that produce the failure modes,
so a local replica does not reproduce them. The transcripts make the gap concrete: the sandbox
ran an older version than production, the agent could not query the sandbox database directly,
external calls left transactions idle while holding locks, and deadlocks only emerged under the
production cluster's concurrent load. None of these is something a single local process
reproduces. Fault-injection testbeds such as AIOpsLab \citep{aiopslab} show that this kind
of environment can be built for evaluation; a replay harness, built on traffic capture or a
shadow environment fed by production traffic, applies the same idea to debugging, and is the
missing piece between ``I can read the logs'' and ``I can reproduce the bug.''

\textbf{Shortening the guess--ship--observe loop.} Each wrong hypothesis cost a full deploy
cycle, which is why the false starts in Section~\ref{sec:incidents} were expensive. That cost is
not just token spend and the engineer's time to redo the work: a redeploy is slow here. The
delivery pipeline runs roughly twelve minutes end to end across about twenty stages, and the
unit-test stage alone takes about five and a half minutes, nearly half the pipeline. The
mutation-testing stage, before it was disabled, added roughly an hour, about five times the rest
of the pipeline combined. Fast fixes matter at this project's scale of about 3.3~million
messages and hundreds of thousands of customers, since a slow fix leaves customers unserved while
the queue backs up. Tooling that lets an agent test a hypothesis against production before
deploying, such as dry-run execution, \texttt{EXPLAIN} against a live replica, or a fast canary
with fine-grained metrics, collapses the loop from ``ship and watch'' to ``ask and answer.''

\textbf{Externalizing the hypothesis.} The retract-and-retry pattern, a production analogue of
the repetitive cycles observed in failing benchmark trajectories \citep{bouzenia2025trajectories}, suggests
that agents should be forced to record each hypothesis and the evidence that retired it, rather
than reasoning privately and correcting silently. The repository's memory of refuted hypotheses
already did this and is the cheapest intervention of all: a hypothesis ledger makes the
false-start chain auditable and, over time, teaches the agent which classes of database failure
to stop guessing about.

Finally, these fixes are as much about the agent's harness (the loop that grants tools,
permissions, and context) as about the database. Harnesses today run a debugging task through the
same loop as a code-generation task. A harness tuned for low-access debugging would treat a
hypothesis as a first-class object, persisted with its evidence across turns, and would prefer
richer, safer read-only primitives, such as a forensics tool rather than raw log search.

\section{Limitations}\label{sec:limitations}

This study is a retrospective field observation of a single agent on a single system, and its
limitations follow from that design.

\emph{Generalizability.} The findings come from one message-dispatch engine with unusually heavy
locking and concurrency, one team, and one agent, over a ten-day incident window. Whether the
pattern holds for other systems, teams, or agents is open.

\emph{Causality and model capability.} We observed access gaps and false starts together, but
without a controlled comparison we cannot show that the former caused the latter. The same
limitation applies to model capability. Most calls were served by an open-weights model, we did
not attribute individual hypotheses to the model that produced them, and the model mix reflects
configuration choices made during ordinary work rather than a deliberate comparison. A frontier
model might have avoided some of the reasoning errors in the mixed incidents.

\emph{Selection.} The eight incidents are those of a single ten-day window, and only those the
team worked on with the agent. With so few cases, the association between concurrency and false
starts holds for our sample only.

\emph{Construct validity and coding.} The constructs, namely the false start, its source
(evidence-consistent or fabrication), and the reasoning $\times$ access matrix, are our
retrospective interpretation of eight incidents. A single author coded them, and that author was
also the participant, so the matrix is a derived taxonomy rather than a validated measurement; we
did not compute inter-rater agreement or perform member checking with the other developers.
Quotations from Portuguese turns are the author's translations.

\emph{Confounds.} The codebase was first built with an AI coding agent and inherited defects from
that phase, so the density of concurrency bugs may be higher than in a conventionally developed
system. A human ran the agent's SQL through the read-only proxy. Over the ten-day window the
agent wrote 28 such queries---25 of them in the single race incident---so this access was real
but concentrated; we did not measure its round-trip latency.

\section{Related Work}\label{sec:related}

Prior work on LLM agents for operations and debugging differs from ours along three axes: what
evidence the agent is given, where the agent runs, and what counts as a failure. We organize the
discussion around these axes rather than around individual systems.

\emph{Evidence: curated bundles versus interrogation.} Most incident-management systems assume
the diagnostic evidence has already been collected, and they place the LLM at the end of that
pipeline. RCACopilot \citep{rcacopilot} runs predefined incident handlers that gather diagnostic
data, then asks the LLM to predict a root-cause category from that bundle and from similar
historical incidents. The design is effective precisely because the expensive step, deciding what
to collect, is encoded by engineers in advance. Roy et al.\ \citep{roy2024rca} move one step toward
autonomy: a ReAct agent retrieves documents and, in a case study, calls the team's own diagnostic
services. RCAgent \citep{rcagent} reaches a similar design from a different constraint: its
tool-augmented agent runs on internally deployed models because company data cannot leave the
organization. In all three systems the agent chooses among, or reads from, evidence channels that
someone already built and authorized. In our setting no such channels existed for the decisive
evidence: the lock graph, the full deadlock \texttt{DETAIL}, and the live row state. The agent's
false starts arose in that gap. Our contribution is to make that missing channel, rather than the
reasoning step that consumes it, the object of study.

\emph{Databases and privileged telemetry.} Database-specific agents make the same assumption,
more strongly. D-Bot \citep{dbot} extracts diagnosis knowledge from documentation and coordinates
several expert agents through tree-search reasoning over system metrics and statistics views.
Panda \citep{panda} argues that ungrounded LLM answers are too generic for database
engineers; it adds grounding in runtime context, verification through cited evidence, and an
affordance component that flags the risk of a proposed action. Our observations support Panda's
diagnosis of the problem: our agent's hypotheses were reasonable but ungrounded in runtime lock
state. Both systems, however, treat access to \texttt{pg\_stat\_activity}, \texttt{pg\_locks},
and metric histories as a given, which holds for a DBA tool running beside the database. Our
agent worked as an application developer's assistant, separated from those views by a VPN, an
identity layer, and a read-only proxy that a human had to operate. In D-Bot's and Panda's terms,
our incidents are cases where the grounding layer was simply absent. Our forensics-API proposal
(Section~\ref{sec:tooling}) can be read as the minimal, security-compatible slice of that layer
that an application-side agent needs.

\emph{Environment: benchmarks versus production.} A recent line of benchmarks evaluates
operations agents in controlled environments. AIOpsLab \citep{aiopslab} deploys microservice
applications, injects faults, generates workloads, and exposes telemetry through a uniform
agent--cloud interface. ITBench \citep{itbench} spans site-reliability, compliance, and
FinOps scenarios, and reports that state-of-the-art agents resolve only a small fraction of the
SRE tasks. OpenRCA \citep{openrca} asks agents to locate root causes in 335 failures from
enterprise systems, with over 68~GB of logs, metrics, and traces. The best model solves only
about 11\% of cases. These low scores are often read as evidence of a reasoning ceiling. Our
field data suggest a complementary reading. Benchmarks give the agent complete observability by
construction: every injected fault leaves its telemetry in the bundle. Their failures therefore
isolate the cost of navigating abundant evidence. In production, the evidence for a concurrency
failure may never be recorded at all, so the cost is incurred earlier, at collection time.
Neither setting captures the other's failure mode, which is why we argue that benchmarks should
also vary what the agent can see, not only how hard the fault is to explain.

\emph{Failure: final answers versus trajectories.} Coding-agent research has shifted from
end-to-end resolution rates toward the structure of agent behavior. SWE-bench \citep{swebench}
established issue resolution as the standard task. SWE-agent \citep{sweagent} showed that the
design of the agent--computer interface, meaning the tools and the feedback they return,
substantially changes performance with the model held fixed. Agentless \citep{agentless} showed
that a simple localize--repair--validate pipeline can match or outperform more autonomous
agents. Bouzenia and Pradel \citep{bouzenia2025trajectories} analyze thought--action--result trajectories and
find that failing runs tend to fall into repetitive cycles, whereas successful runs balance
exploration with validation. Our false starts are a production analogue of these cycles, with one
important difference: in a benchmark, a failed validation costs a test run, whereas in our
setting the only validation available was a deploy followed by observation of live traffic, which
multiplied the cost of each cycle by the length of the delivery pipeline. Field studies of
agentic coding, such as Watanabe et al.\ \citep{watanabe2026agentic} on Claude Code pull requests, have
so far focused on code-producing tasks and their acceptance. To our knowledge, the diagnosis of a
running system has not been studied with the agent's own transcripts as evidence.

\emph{Reasoning about concurrency and the hallucination framing.} Our claim that reasoning was
not the dominant bottleneck is consistent with evidence that LLMs can reason about concurrency
when the relevant facts are in view. Jain and Purandare \citep{jain2026concurrency} find that frontier
models reliably detect data races and deadlocks in concurrent programs under sequential
consistency. That task gives the model a small program that fits in its context, with the bug
visible in the code. Our agent could read the whole repository, but at about 406{,}000 lines it
had to choose what to read, and the interleaving that actually occurred was recorded only in
runtime state. Finally, Zhang et al.\ \citep{zhang2025hallucination} show that many of the hallucinations in
repository-level code generation stem from missing context, such as unknown APIs or
project-specific facts. We draw the same line between fabrication and missing information, but at
runtime rather than in code. Within the incidents, our agent's wrong hypotheses were not
fabrications; they were explanations consistent with the evidence it could see. Within the
standard architecture of LLM-based agents (planning, memory, tools, action) \citep{agentsurvey},
all of these strands point to the tool and perception layer, at least as much as to the
reasoning core, as the place where production-debugging agents currently fail.

\section{Conclusion}\label{sec:conclusion}

We asked what limits an LLM coding agent's diagnosis of production database incidents: its
reasoning or its access. Across eight incidents and twenty-two fix commits, the answer was mostly
the latter. The agent's diagnostic reasoning was largely sound: it formed plausible hypotheses,
noticed its own contradictions, and retired wrong hypotheses through measurement. Yet every
incident whose root cause depended on runtime concurrency began with a false start. The agent
had the entire codebase, yet for these failures the code was not enough. In most of them, the
evidence that would have settled the diagnosis existed but was out of reach: the live lock view,
the full deadlock report, the row-level state, and the environment in which the failure
appeared. Where the agent did reason badly, by anchoring early, missing a configuration value, or
misreading rollback semantics, the error occurred alongside an access gap that delayed its
correction. The reasoning $\times$ access matrix makes the pattern explicit: every
concurrency-dependent incident had an access gap, and no incident was limited by reasoning alone.

This finding changes where the next investment should go. For tool builders, the
highest-leverage work is not a larger model but a forensics layer. It should be a read-only,
audited interrogation surface, a recorder that preserves transient lock and deadlock state, and
a way to reproduce production's concurrent behavior. All of it must live inside the security
controls that make production access hard to grant in the first place. For researchers who build
benchmarks, the finding implies that agents should be evaluated under partial observability, and
not only on how hard a fault is to explain. For practitioners, the finding is a caution: the
agent amplified the leverage of an engineer who understood concurrency, but it did not replace
the judgment needed to tell when a plausible hypothesis was wrong. In mission-critical systems,
the first bottleneck of today's debugging agents is access, and it is the one that tooling can
remove.

\emph{Future work.} The most direct next step is to turn our observation into a controlled
comparison, crossing model capability with access. Replaying the same incidents with and without
a forensics layer, while holding the model fixed, would test whether access causes the false
starts, a claim this study can only suggest. The comparison would also measure what each wrong
hypothesis costs in deploy cycles and time to resolution. A second step is to build and evaluate
the components we propose, including a lock-and-deadlock flight recorder for PostgreSQL and a
read-only interrogation API brokered by the existing access layers. Third, our findings come from
one system, one team, and a single coder; multi-site replication across other database engines,
workloads, and agents is needed, with independent coders and inter-rater agreement, to establish
whether the ``access more than reasoning'' pattern generalizes and to refine the matrix into a
validated taxonomy. Finally, we see room for benchmarks that vary the agent's observability
explicitly, for example by truncating logs or withholding lock views in testbeds such as
AIOpsLab \citep{aiopslab} or OpenRCA \citep{openrca}. Such benchmarks would let the
community separate failures of reasoning from failures of access at scale, which our single
field study could not do.

\section*{Data Availability}
The raw data (session transcripts, commit history, and production metrics) cannot be shared,
because they contain the company's proprietary information.

\section*{Declaration of Competing Interest}
The author was employed by the company under study during the period covered by the data and
directed the agent as part of that employment. The author declares no other competing interests.

\section*{Declaration of Generative AI and AI-Assisted Technologies in the Manuscript Preparation Process}
During the preparation of this work the author used Claude Code (an AI coding agent) to support
the analysis of the session transcripts and repository data and to assist with drafting the
manuscript. Because the object of study is the same tool, its use is also discussed as a threat
to validity in Section~\ref{sec:limitations}. After using this tool, the author reviewed and
edited the content as needed and takes full responsibility for the content of the published
article.

\bibliographystyle{cas-model2-names}
\bibliography{refs}

\end{document}